\documentclass[12pt]{aipproc}
\layoutstyle{6x9}

\catcode`@=12
\def\bea{\begin{eqnarray}}
\def\eea{\end{eqnarray}}
\def\be{\begin{equation}}
\def\ee{\end{equation}}
\begin{document}
\begin{flushright}
\end{flushright}
\vspace{0.5in}
\title{TeV scale model for neutrino masses, dark matter and leptogenesis}
\author{Narendra Sahu} {address={Theory Division, 
Physical Research Laboratory, Navarangpura, Ahmedabad, 380 009, 
India},}

\begin{abstract}
We present a TeV scale model for leptogenesis where the origin 
of neutrino masses are independent of the scale of leptogenesis. 
As a result, the model could be extended to explain {\it dark 
matter, neutrino masses and leptogenesis at the TeV scale}. The 
most attractive feature of this model is that it predicts a few 
hundred GeV triplet Higgs scalar that can be tested at LHC or ILC.    

\end{abstract}
\maketitle

In the type-I seesaw models~\cite{canonical_seesaw} the physical 
neutrino masses are largely suppressed by the scale of lepton (L) 
number violation, which is also the scale of leptogenesis~\cite{
fukugita.86}. The observed baryon (B) asymmetry, defined by 
$(n_B/n_\gamma)_0 = (6.15\pm 0.25) \times 10^{-10}$, and the low 
energy neutrino oscillation data combinely then give a lower bound 
on the scale of leptogenesis to be $\sim {\mathcal O}(10^9)$ 
GeV~\cite{di_bound}. Alternately in the type-II seesaw 
models~\cite{tripletseesaw} it is equally difficult to generate 
$L$-asymmetry at the TeV scale because the interaction of $SU(2)_L$ 
triplets with the gauge bosons keep them in equilibrium up to 
a very high energy scale $\sim {\mathcal O}(10^{10})$ GeV~\cite{
ma&sarkar_prl}. Thus irrespective of the seesaw models, the scale of 
leptogenesis can not be less than $\sim {\mathcal O}(10^9)$ GeV~\cite{
sahu&sarkar_prd}. This is because in these class of models the $L$-number 
violation required for neutrino masses and leptogenesis is same.  

In a previous work~\cite{sahu&sarkar.07} we proposed a new mechanism 
of leptogenesis at the TeV scale. We ensure that the lepton number 
violation required for the neutrino masses does not conflict with 
the lepton number violation required for leptogenesis. As a result the 
model could be extended to explain dark matter, neutrino masses and 
leptogenesis at the TeV scale. Moreover, the model predicts a few hundred 
GeV triplet Higgs scalar whose decay through the same sign dilepton signal 
could be tested at LHC or ILC. 

{\bf The Model:} We now describe the salient features of the model. 
In addition to the quarks, leptons and the usual Higgs doublet 
$\phi \equiv (1,2,1)$, we introduce two triplet Higgs scalars 
$\xi \equiv (1,3,2)$ and $\Delta \equiv (1,3,2)$, two singlet 
scalars $\eta^- \equiv (1,1,-2)$ and $T^0 \equiv (1,1,0)$, and a 
doublet Higgs $\chi\equiv (1,2,1)$. The transformations of the fields 
are given under the standard model (SM) gauge group $SU(3)_c \times 
SU(2)_L \times U(1)_Y$. There are also three heavy singlet fermions 
$S_a \equiv (1,1,0), a = 1,2,3$. A global symmetry $U(1)_X$ allows us 
to distinguish between the $L$-number violation required for neutrino 
masses and the $L$-number violation required for leptogenesis. Under 
$U(1)_X$ the fields $\ell_{iL}^T \equiv (\nu, e)_{iL} \equiv (1,2,-1)$, 
$e_{iR} \equiv (1,1,-2)$, $\eta^-$ and $T^0$ carry a quantum number 1, 
$\Delta$, $S_a$, $a=1,2,3$ and $\phi$ carry a quantum number zero, while 
$\xi$ and $\chi$ carry quantum numbers -2 and 2 respectively. We assume 
that $M_\xi \ll M_\Delta$. However, as we shall see later, they contribute 
equally to the neutrino masses.

We now write down the Lagrangian symmetric under $U(1)_X$. The terms in 
the Lagrangian, relevant to the rest of our discussions, are given by
\bea
-{\cal L } &\supseteq & f_{ij} \xi \ell_{iL} \ell_{jL} + \mu \Delta^\dagger
\phi \phi + M_\xi^2  \xi^\dagger \xi + M_\Delta^2 \Delta^\dagger
\Delta + h_{ia} \bar e_{iR} S_a \eta^- + M_{sab} S_a S_b + y_{ij}\phi
\bar \ell_{iL} e_{jR} \nonumber\\
&&+ M_T^2 T^\dagger T + \lambda_T |T|^4 + \lambda_\phi |T|^2 |\phi|^2 
+ \lambda_\chi |T|^2 |\chi|^2 + f_T \xi \Delta^\dagger T T\nonumber\\
&& + \lambda_{\eta \phi}|\eta^-|^2 |\phi|^2 + \lambda_{\eta \chi}
|\eta^-|^2|\chi|^2 + V_{\phi \chi} + h.c.\,,
\label{lagrangian}
\eea
where $V_{\phi \chi}$ constitutes all possible quadratic and quartic 
terms symmetric under $U(1)_X$. We introduce the $U(1)_X$ symmetry 
breaking soft terms
\begin{equation}
-\mathcal{L}_{soft} = m_T^2 T T + m_\eta \eta^- \phi \chi + h.c. \,.
\label{soft-terms}
\end{equation} 
If $T$ carries the $L$-number by one unit then the first term 
explicitly breaks $L$-number in the scalar sector. The second term 
on the other hand conserves $L$-number if $\eta^-$ and $\chi$ possess 
equal and opposite $L$-number. This leads to the interactions of the 
fields $S_a, i=1,2,3$ to be $L$-number conserving. As we shall discuss 
later, this can generate the $L$-asymmetry of the universe, while the 
neutrino masses come from the $L$-number conserving interaction term 
$\xi \Delta^\dagger T T $ after the field $T$ acquires a vacuum expectation 
value (VEV).

At tree level the Higgs field $\Delta$ acquires a very small VEV 
\begin{equation}
\langle \Delta \rangle = -\mu {v^2 \over M_\Delta^2}\,,
\end{equation}
where $v = \langle  \phi \rangle$, $\phi$ being the SM Higgs doublet. 
However, we note that the field $\xi$ does not acquire a VEV at the 
tree level. 

At a few TeV the scalar field $T$ acquires a VEV. This gives rise to 
a mixing between $\Delta$ and $\xi$ through the effective mass term
\begin{equation}
-{\cal L }_{\Delta \xi} = m_{s}^2 \Delta^\dagger \xi ,
\end{equation}
where the mass parameter $m_s = \sqrt{f_T \langle T \rangle^2}$ is of 
the order of TeV, similar to the mass scale of $T$. The effective 
couplings of the different triplet Higgs scalars, which give the 
$L$-number violating interactions for neutrino masses, are 
then given by
\begin{equation}
-{\cal L }_{\nu - mass} = f_{ij} \xi \ell_i \ell_j + \mu {m_s^2 \over 
M_\Delta^2} \xi^\dagger \phi \phi + f_{ij} {m_s^2 \over M_\xi^2} 
\Delta \ell_i \ell_j + \mu \Delta^\dagger \phi \phi + h.c.\,.
\label{flavour_vio}
\end{equation}
The field $\xi$ then acquires an induced VEV,
\be
\langle \xi \rangle = -\mu {v^2 m_s^2 \over M_\xi^2 M_\Delta^2}\,.
\label{xi_vev}
\ee
The VEVs of both the fields $\xi$ and $\Delta$ will contribute to
neutrino masses by equal amount and thus the neutrino masses are given 
by
\begin{equation}
( {m}_{\nu} )_{ij}= - f_{ij} \mu {v^2 m_s^2 \over M_\xi^2 M_\Delta^2}\,.
\label{neutrino_mass}
\end{equation}
Assuming that $M_\Delta \sim \mu \simeq {\mathcal O}(10^{15})$ GeV 
one can find $M_\xi$ to be of the order of a few hundred GeV. This 
makes the model predictive since the decay of $\xi$ through the 
same sign dilepton can be verified in the near future colliders 
(LHC/ILC). 

{\bf Leptogenesis:} Since the absorptive part of the off-diagonal 
one loop self energy terms in the decay of triplets $\Delta$ and 
$\xi$ is zero, their decay can't produce any $L$-asymmetry even 
though their decay violate $L$-number. However, the possibility 
of erasing any pre-existing $L$-asymmetry through the 
$\Delta L=2$ processes mediated by $\Delta$ and $\xi$ should not 
be avoided. In particular, the important erasure processes are: $\ell \ell 
\leftrightarrow \xi \leftrightarrow \phi \phi $ and $\ell \ell 
\leftrightarrow \Delta \leftrightarrow \phi \phi$. If $m_s^2\ll 
M_\Delta^2$ then the $L$-number violating processes mediated 
through $\Delta$ and $\xi$ are suppressed by $(m_s^2/M_\xi^2 M_\Delta^2)$ 
and hence practically don't contribute to the above erasure processes. 
Thus a fresh $L$-asymmetry can be produced at the TeV scale. 

Without loss of generality we shall work in a basis in which $M_{sab}$ 
is diagonal and $M_3 > M_2 > M_1$, where $M_a = M_{saa}$. In this basis, 
the decay of the singlet fermions $S_a$, $a=1,2,3$ generates an equal 
and opposite $L$- asymmetry between $e_{iR}^-$ and $\eta^+$ fields 
through 
\begin{eqnarray}
S_a &\to & e_{iR}^- + \eta^+ \nonumber \\
& \to & e_{iR}^+ + \eta^- \,,\nonumber
\end{eqnarray}
since $e_{iR}^-$ and $\eta^+$ carry equal and opposite $L$- number.
The one-loop self-energy and vertex-type diagrams that can interfere 
with the tree-level decays to generate a CP-asymmetry
\begin{equation}
\epsilon = - \sum_i \left[ {\Gamma(S_1 \to e_{iR}^- \eta^+)
- \Gamma(S_1 \to e_{iR}^+ \eta^-) \over \Gamma_{tot} (S_1)}
\right] \simeq  {1 \over 8 \pi} {M_1 \over M_2} {{\rm Im} [
(h h^\dagger)_{i1}^2] \over \sum_a |h_{a1}|^2}\,.
\label{cp_asymmetry}
\end{equation}
If these two asymmetries cancel with each other then there should not 
be any left behind $L$-asymmetry. However, as we see from the Lagrangians 
(\ref{lagrangian}) and (\ref{soft-terms}) that none of the interactions 
that can transfer the $L$-asymmetry from $\eta^-$ to the lepton doublets 
while $e_R$ is transferring the $L$-asymmetry from the singlet sector to 
the usual lepton doublets through $\phi \bar{\ell}_L e_R$ coupling. Note 
that the coupling, through which the asymmetry between $\eta^-$ and 
$e_R^+$ produced, is already gone out of thermal equilibrium. 
So, it will no more allow the two asymmetries to cancel with 
each other. The asymmetry in the $\eta$ fields is finally transferred 
to the $\chi$ fields through the trilinear soft term introduced in 
Eq. (\ref{soft-terms}). The $B+L$ violating sphaleron processes then 
convert the $L$- asymmetry in the lepton doublets to a net $B$- asymmetry. 
Since the source of $L$-number violation for this 
asymmetry is different from the neutrino masses, there is no bound on 
the mass scale of $S_1$ from the low energy neutrino oscillation data. 
Therefore, the mass scale of $S_1$ can be as low as a few TeV.

{\bf Dark matter:} As the universe expands the temperature of the 
thermal bath falls. As a result, below their mass scales, the heavy 
fields $\eta^-$ and $T^0$ are annihilated to the lighter fields 
$\phi$ and $\chi$. Notice that there is a surviving $Z_2$ symmetry 
of the Lagrangians (\ref{lagrangian}) and (\ref{soft-terms}) under 
which $S_a, a=1,2,3$, $\eta^-$ and $\chi$ are odd while all other 
fields are even. Since the neutral component of $\chi$ is the 
lightest one it can be stable because of $Z_2$ symmetry. Therefore, 
the neutral component of $\chi$, having mass in a range 50 GeV - 100 GeV, 
behaves as a dark matter. 

{\bf Collider Signature:} The doubly charged component of the light 
triplet Higgs $\xi$ can be observed through its decay into same sign 
dileptons~\cite{collider_signature}. Since $M_\Delta \gg M_\xi$, the 
production of $\Delta$ particles in comparison to $\xi$ are highly 
suppressed. Hence it is worth looking for the signature of 
$\xi^{\pm\pm}$ either at LHC or ILC. Once it is produced, $\xi$ mostly 
decay through the same sign dileptons: $\xi^{\pm \pm} \rightarrow 
\ell^\pm \ell^\pm$. Note that the doubly charged particles can not 
couple to SM quarks and therefore the SM background of the process 
$\xi^{\pm\pm}\rightarrow \ell^\pm \ell^\pm$ is quite clean and hence 
the detection will be unmistakable.

\bibliographystyle{unsrt}

\end{document}